\documentclass[conference]{IEEEtran}
\IEEEoverridecommandlockouts
\usepackage{cite}
\usepackage{booktabs}
\usepackage{amsmath,amssymb,amsfonts}
\usepackage{algorithmic}
\usepackage{graphicx}
\usepackage{textcomp}
\usepackage{xcolor}
\usepackage{subfigure}
\usepackage[draft]{hyperref}
\usepackage{dblfloatfix} 

\usepackage{tikz}
\usetikzlibrary{arrows,positioning,shapes.geometric, calc}

\def\BibTeX{{\rm B\kern-.05em{\sc i\kern-.025em b}\kern-.08em
    T\kern-.1667em\lower.7ex\hbox{E}\kern-.125emX}}
\usepackage{orcidlink}

\usepackage{siunitx}
\DeclareSIUnit{\belmilliwatt}{Bm}
\DeclareSIUnit{\dBm}{\deci\belmilliwatt}

\DeclareMathOperator{\sinc}{sinc}

\IEEEoverridecommandlockouts
\IEEEpubid{\makebox[\columnwidth]{978-1-6654-0579-9/22/\$31.00~\copyright{}2022 IEEE \hfill} \hspace{\columnsep}\makebox[\columnwidth]{ }}
    
\begin{document}

\title{Enabling Radio Sensing for Multimodal\\ Intelligent Transportation Systems: \\From Virtual Testing to Immersive Testbeds\\

}

\author{\IEEEauthorblockN{Paul Schwarzbach\orcidlink{0000-0002-1091-782X}, Jonas Ninnemann\orcidlink{0000-0001-7988-079X}, and Oliver Michler\orcidlink{0000-0002-8599-5304}}
\IEEEauthorblockA{\textit{Chair of Transport Systems Information Technology} \\
\textit{Institute of Traffic Telematics} \\
\textit{Technische Universit\"at Dresden}\\
Dresden, Germany \\
\{paul.schwarzbach, jonas.ninnemann, oliver.michler\}@tu-dresden.de}
}

\maketitle

\begin{abstract}
In this paper, the necessity for application-oriented development and evaluation of Joint Communication and Sensing (JC\&S) applications, especially in transportation, is addressed. More specifically, an integrative evaluation chain for immersively testing JC\&S location capabilities, reaching from early-stage testing, over model- and scenario-enabled ray tracing simulation, to real-world evaluation (laboratory and field testing) is presented. This includes a discussion of both challenges and requirements for location-aware applications in Intelligent Transportation Systems. Within this scope, a reproducible methodology for testing sensing and localization capabilities is derived and application scenarios are presented. This includes a proposal of a scenario-based sensing evaluation using radio propagation simulation. The paper empirically discusses a proof-of-concept of the developed method given a smart parking scenario, in which a passive occupancy detection of vehicles is performed. The conducted findings underline the need for scenario-based JC\&S evaluation in both virtual and real-world environments and proposes consecutive research work.
\end{abstract}


\section{Introduction}
\label{sec:Intro}
Mobility demands are a fundamental necessity in all humans daily lives. However, they need to be met with a minimum of traffic volume. Driven by ecological, demographic and also technological changes, constraints on adaptation in the transportation sector are desperately required \cite{zhang_survey_data_driven_its_2011}. In research, technology-aided Intelligent Transportation Systems (ITS) provide comprehensive state monitoring, intelligent networking and holistic control of transportation components \cite{Papadimitratos_its_technologies_applications_outlook_2009}. Enhanced by the capabilities of the 5G technology, transportation efficiency and safety will be enhanced \cite{5G-ITS}. Furthermore, a key-enabler for ITS is location-awareness. However, its performance demands and sensing capabilities most likely will not be met with solely 5G. To overcome this shortcoming, Joint Communication and Sensing (JC\&S) is one of the core features of emerging Beyond 5G (B5G) radio technologies \cite{bourdoux_6g_2020-3}, which have recently been addressed \cite{ziegler_stratification_2019, aazhang_key_2019}.

\begin{figure}[ht]
    \centering
    \includegraphics[width=0.76\linewidth]{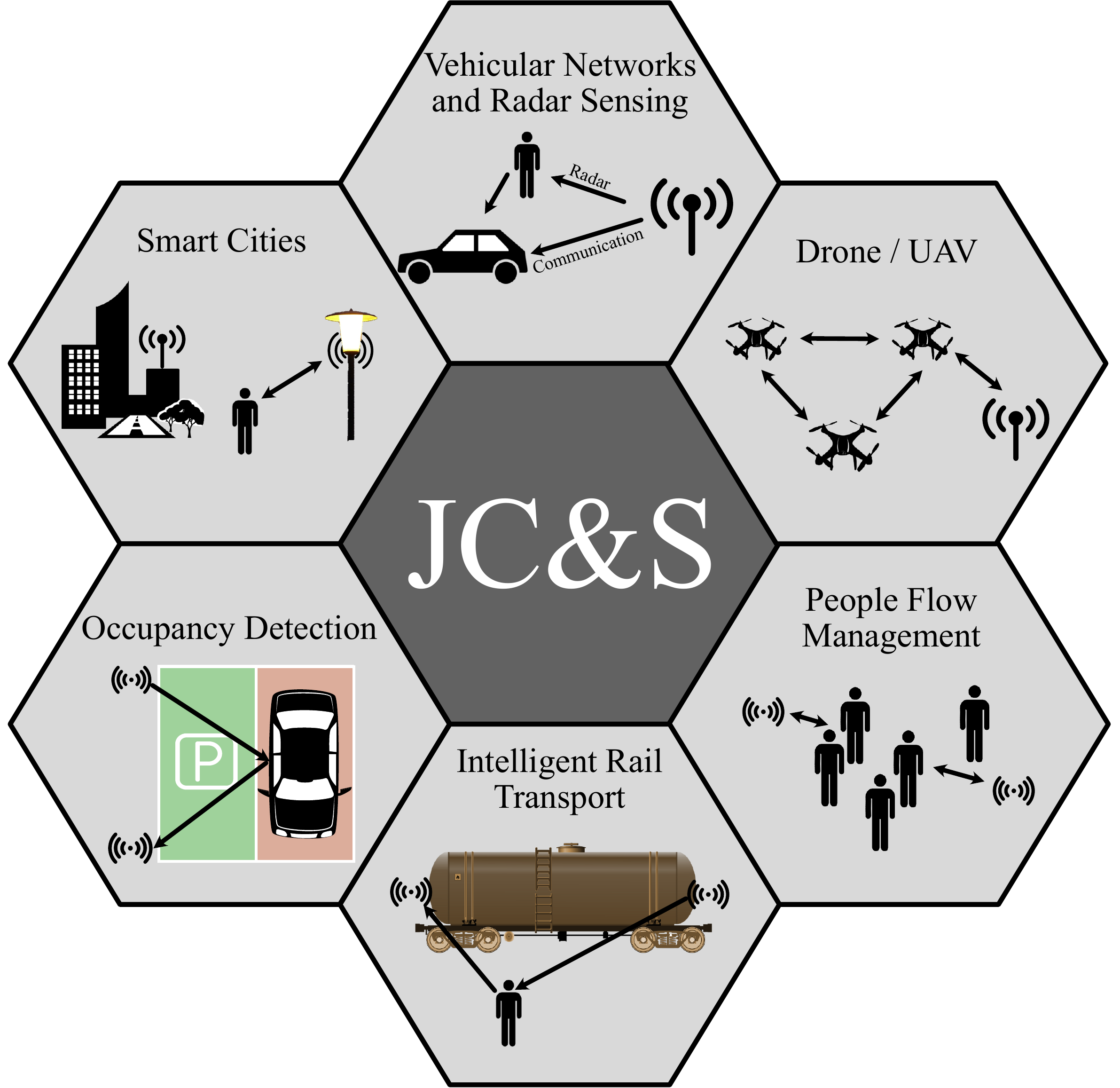}
    \caption{Location-aware JC\&S applications in ITS.}
    \label{fig:GA}
\end{figure}

Potential techniques, features, architectures, improvements and new applications towards B5G are discussed in \cite{dang_what_2020-1, yang_6g_2019, zhang_6g_2019}. With the physical layer innovations planned for B5G systems, accompanying benefits for localization, sensing and mapping applications also arise. These include: network densification, larger bandwidths and higher frequency bands. While future improvements in availability and accuracy for location-based services and context-awareness can potentially improve manifold applications, new location-enabled capabilities are especially required for transportation and mobility demands. Fig.~\ref{fig:GA} depicts a brief overview of JC\&S application scenarios, whereas corresponding use cases and accuracy requirements are provided in Tab.~\ref{tab:Applications}.

With arising challenges like automation, densification and inter-modality, seamless and integrated systems are imperatively required. However, with these demands, complexity for usability simultaneously increases. In contrary, current research in the field of B5G addresses visions, proof-of-concepts or detailed contributions of implementation and design aspects, with hardly any focus on scenario-based sensing evaluation.

\begin{table}[ht]
\caption{Applications, use cases and accuracy requirements of JC\&S in multimodal ITS.}
\begin{center}
\begin{tabular}{p{1.4cm} p{3.3cm} p{2.7cm}}
\toprule
\textbf{Applications} & \textbf{Use Cases} & \textbf{Requirements}\\
\toprule
Automotive &
    Autonomous driving & $<\SI{0.29}{\meter}$ \cite{reid_localization_for_autonomous_vehicles_2019} \\
   \cite{cao_joint_2020, zhang_enabling_2021-1} & JC\&S vehicular networks & $<\SI{1.10}{\meter}$ \cite{ansari_DSRC_analysis_2017} \\
    & Smart Parking surveillance & Area-selective $<\SI{1}{\meter}$\\
\midrule
Smart Cities  & 
    Cooperative GLOSA & $<\SI{1}{\meter}$  \cite{etsi_ts_122_261} \\
   \cite{zhang_enabling_2021-1} & Real-time monitoring & $1 - \SI{10}{\meter}$ \cite{etsi_ts_122_261} \\ 
\midrule
Rail & 
    Passive safety systems & $< \SI{1}{\meter}$\\
\midrule
UAVs \cite{bourdoux_6g_2020-3} & 
    Tracking and surveillance & $<\SI{0.5}{\meter}$ \cite{xinlie_urllc_use_cases_2019}\\
\midrule
   Connected & Automated Safety Checks & Area-selective \cite{Cogalan_inflight_connectivity_2018} \\
   Cabin & Passenger monitoring & Seat-selective \cite{Schultz_Cabin_OR_2019,Schwarzbach_covid_boarding_2020}
   \\
\midrule
Industrial &
    Asset Tracking & $0.3 - \SI{1}{\meter}$  \cite{etsi_ts_122_261} \\
    IoT & Automated Guided Vehicles & $<\SI{0.5}{\meter}$ \cite{xinlie_urllc_use_cases_2019}\\
\bottomrule
\end{tabular}
\label{tab:Applications}
\vspace{-.5cm}
\end{center}
\end{table}

\begin{figure*}[b]
    \centering
    \includegraphics[width=.75\textwidth]{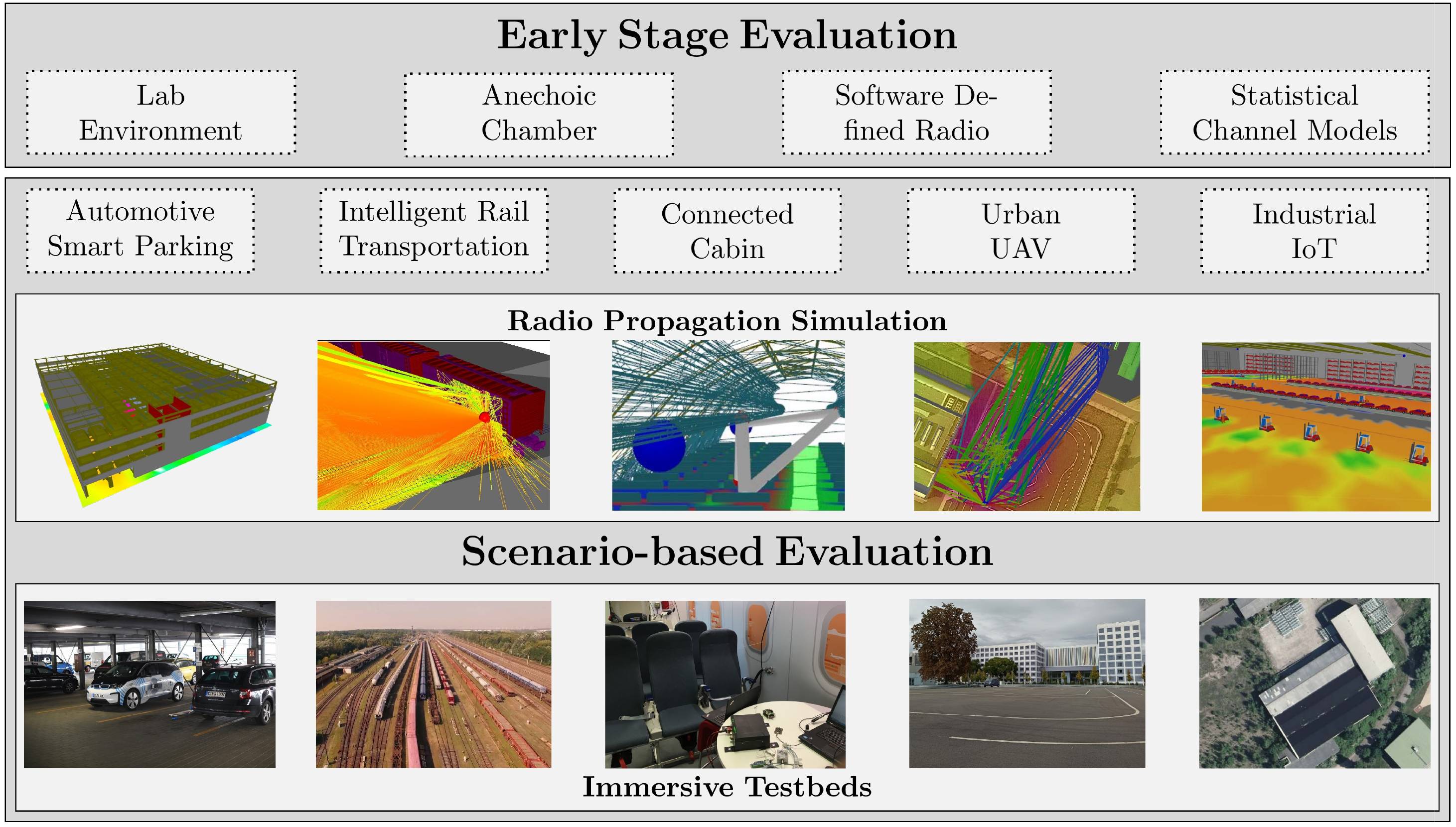}
    \caption{Evaluation stages for JC\&S in ITS scenarios, including early stage testing, radio propagation simulation and testbeds.}
    \label{fig:evaluation}
\end{figure*}

For mobile radio evaluation, channel simulation is conventionally realized by applying consolidated statistical channel models \cite{liu_information-theoretic_2021}, which describe predefined scenarios and corresponding channel statistics, inter alia Line-of-Sight probability and path loss \cite{etsi_tr_38_901}. While originally designed to generate inputs for both wireless communication and active radio localization, input data for sensing can hypothetically also be derived, as published works on multipath-enabled radio sensing \cite{sensors_DFPL, mampi-uwbmultipath-assisted_2020-2} essentially use the channel coefficients, which are also derived from statistical channel models \cite{etsi_tr_38_901}. Though, as pointed out in \cite{wild_joint_2021}, exclusive use of spatial statistical channel models is not suitable for sensing applications. Unlike communication and localization, whose inputs can easily be expressed by means of statistical measures, sensing is entirely dependent on the actual propagation environment. Therefrom, simulation intended for sensing capabilities requires a coherent interdependency of real-world scenarios and their digital twins.


Within the scope of sensing simulation frameworks, the paper suggests and discusses the usage of model-based ray-tracing radio propagation simulation. With these tools available, application-proximate and scenario-enabled evaluation of JC\&S can be achieved, especially in ITS scenarios, where integrated communication system evaluation is generally challenging. This includes the accessibility of scenarios and environments, the availability of technologies and hardware components as well as the adaptability of surveyed data. Fig.~\ref{fig:evaluation} provides a comprehensive overview on possible JC\&S ITS scenarios, including testbed environments and their digital representatives. In addition, common tools for early-stage testing are provided.

Benefits of virtual sensing evaluation include the examination of the interaction and correlation between the scenarios and simulation, aiming to potentially compensate elaborate real-world testing for future scenarios. This is particularly important, as emerging technologies and methods for transportation and multimodal applications are forced within an economic, space-restricted and efficiency driven design due to the immanent mobility of transportation systems. Concluding, while ITS and their performance requirements are demanding for JC\&S systems, the potential benefits for location-awareness, surveillance and information gain are key for efficiency and safety in ITS.

The remainder of this paper is structured as follows: After the introduction (Sec.~\ref{sec:Intro}), a general methodology for the simulation of sensing capabilities in JC\&S is discussed (Sec.~\ref{sec:Methodology}). In Sec.~\ref{sec:framework}, a framework for using the output of radio propagation simulation as the input of a multipath-assisted sensing method is introduced. An evaluation example given an in-house smart parking garage and qualitative sensing results are presented in Sec.~\ref{sec:Results}. The paper concludes with a brief summary and proposals for future work in Sec.~\ref{sec:conclusion}.
 
\section{Problem Formulation}
\label{sec:Methodology}
Evaluation of radio systems has to be performed given the conditions of potential use cases and scenarios. In case of sensing capabilities, the spatial and electromagnetic (EM) properties of the propagation environment needs to be considered and solely statistical channel models are not applicable \cite{wild_joint_2021}. In addition, usable hardware and devices might not be available, especially for conceptual technologies. This generally also applies for exploratory research. 

Therefore, a virtual simulation is well-suited for the task at hand, however, it needs to correlate with the conditions of the scenario. The given use cases can then be digitized and used for simulation. This is achieved by providing a digital model of the propagation environment and a radio model \cite{Zhengqin_ray_tracing_2015}. The former consists of a geometrical model, which comprises the positions and dimensions, as well as the EM properties of all relevant objects within the environment. The radio model includes the physical layer properties of the desired radio technology respectively hardware.

\begin{figure}[h]
    \centering
    \centerline{\includegraphics[width=\linewidth]{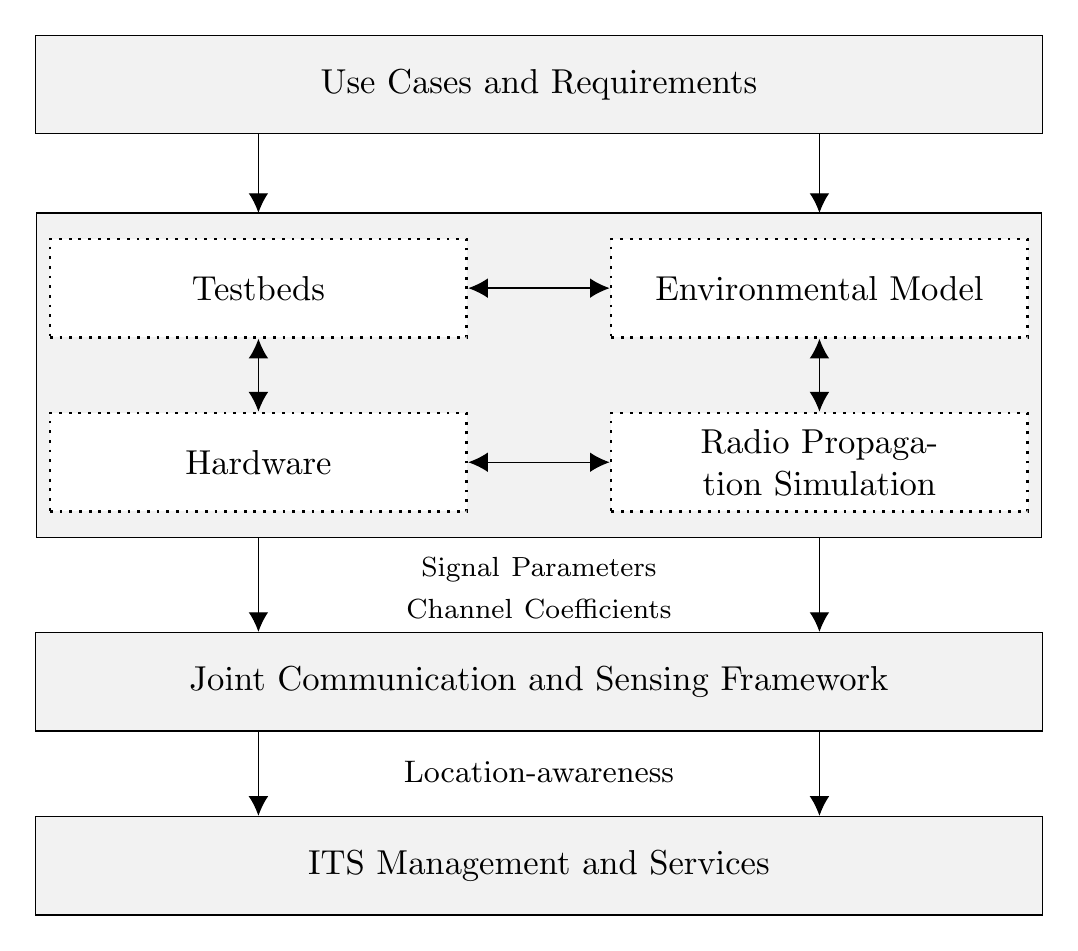}}
    \caption{Methodology of integrative evaluation for JC\&S in ITS.}
    \label{fig:meth}
\end{figure}

Given these boundary conditions, we propose to assess the sensing performance of ITS JC\&S using an integrated and phased evaluation approach, based on initial virtual simulation and real-world environments. A derived methodology, describing the progression from JC\&S use cases towards scenario-based evaluation, is depicted in Fig.~\ref{fig:meth}, where use cases and requirements correspond to the examples in Tab.~\ref{tab:Applications}.

Consequently, both real-world examination and radio simulation provide the relevant signal properties desired for radio sensing. Among others, the Channel Impulse Response (CIR) of the transmission channel provides essential information for passive sensing \cite{sensors_DFPL} and is conventionally provided by ray-tracing simulation. Based on these inputs, location information and surveillance can be derived and used for ITS in order to provide services and controlling.

\section{Virtual Sensing Simulation}
\label{sec:framework}

\begin{figure}[h]
    \centering
    \begin{tikzpicture}[>=latex']
        \tikzset{block2/.style= {draw, shape=rectangle,rounded corners,align=center,text width=2.2cm, minimum width=2.2cm, minimum height=1cm, fill=black!5,},
        block/.style={draw, shape=rectangle,rounded corners,align=center,text width=2cm, minimum width=2cm, minimum height=3.4cm, fill=black!5,},
        }
        
        \begin{scope}[scale=0.75, transform shape]
            \node [block2] (model) {Environmental Model};
            \node [block2, below =0.6cm of model] (radio) {Radio \\ Model};
            \node [block, right of= model, xshift=2cm, yshift=-0.8cm] (simulation) {Deterministic Radio Propagation Simulation};
            \node [block, right =1cm of simulation] (jcas) {Multipath-assisted JC\&S Framework};
            \node [block2, right =0.7cm of jcas] (map) {Heatmap};
            \node [block2, above =0.2cm of map] (position) {Position};
            \node [block2, below =0.2cm of map] (detection) {State Detection};

        \draw[->] (model.east) -- (model.east-|simulation.west);
        \draw[->] (radio.east) -- (radio.east-|simulation.west);
        \draw[->] (simulation.east) -- node [above] {CIR} (jcas.west);
        \draw[<-] (map.west) -- (map.west-|jcas.east);
        \draw[<-] (position.west) -- (position.west-|jcas.east);
        \draw[<-] (detection.west) -- (detection.west-|jcas.east);
        \end{scope}
\end{tikzpicture}
    \vspace{-0.5cm}
    \caption{Data processing flowchart of the virtual sensing simulation, including required inputs and exemplary sensing outputs.}
    \label{fig:flowgraph}
\end{figure}
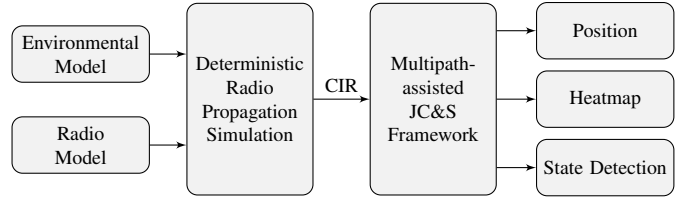

The proposed simulation approach is illustrated in Fig.~\ref{fig:flowgraph} and comprises the following essentials:

\begin{itemize}
    \item Ray-tracing radio propagation simulation and
    \item Multipath-assisted sensing framework.
\end{itemize} 

The former can generally be applied for JC\&S evaluation, whereas the latter is based on previously discussed realizations of multipath-assisted sensing \cite{sensors_DFPL, ninnemann2022multipathassisted} and serves as an exemplary implementation to illustrate the entire toolchain.

\begin{figure*}[b]
\centering
\subfigure[]{
    \includegraphics[width=0.19\textwidth]{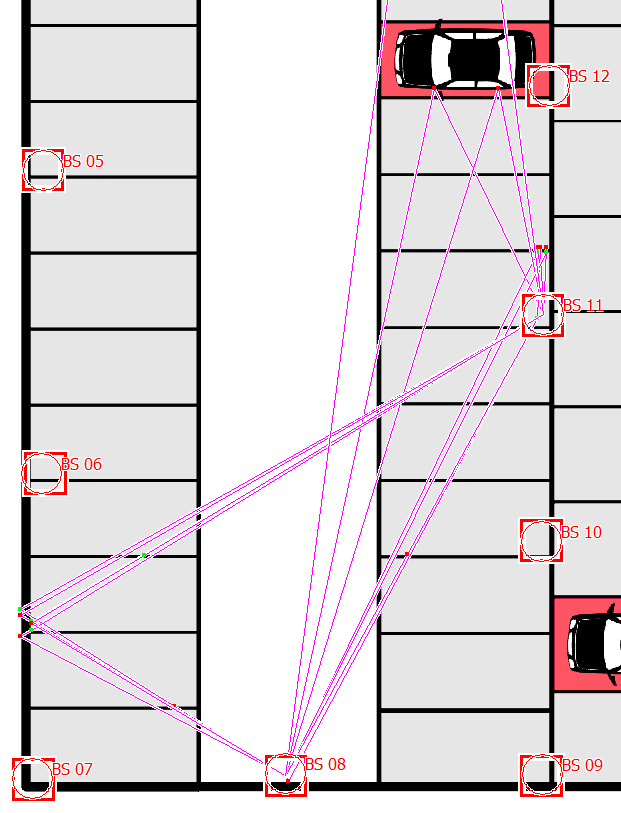}
    \label{fig:propagation_paths}
}
\subfigure[]{
    \includegraphics[width=0.37\textwidth]{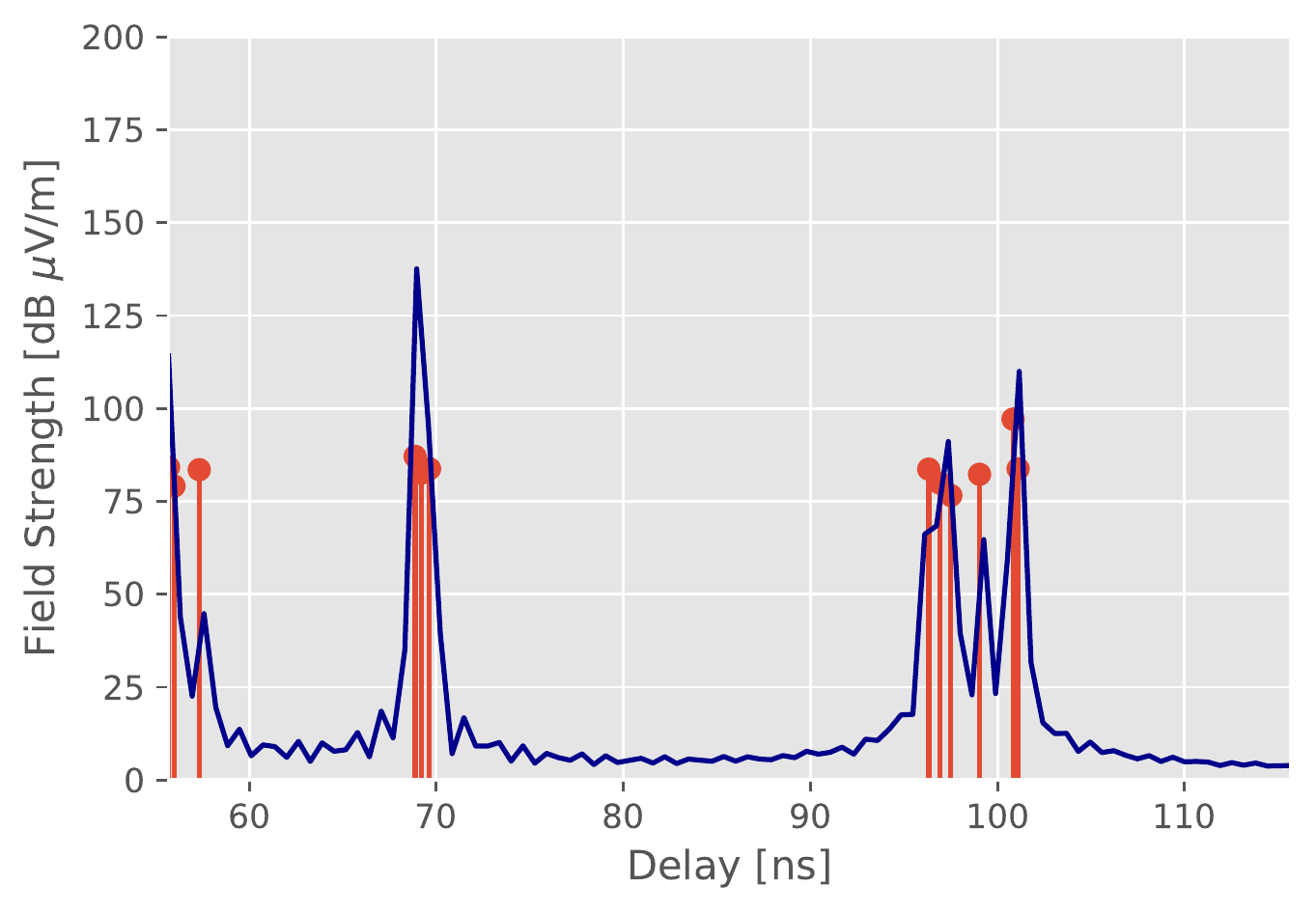}
    \label{fig:cir_bandlimited}
}
\subfigure[]{
    \includegraphics[width=0.37\textwidth]{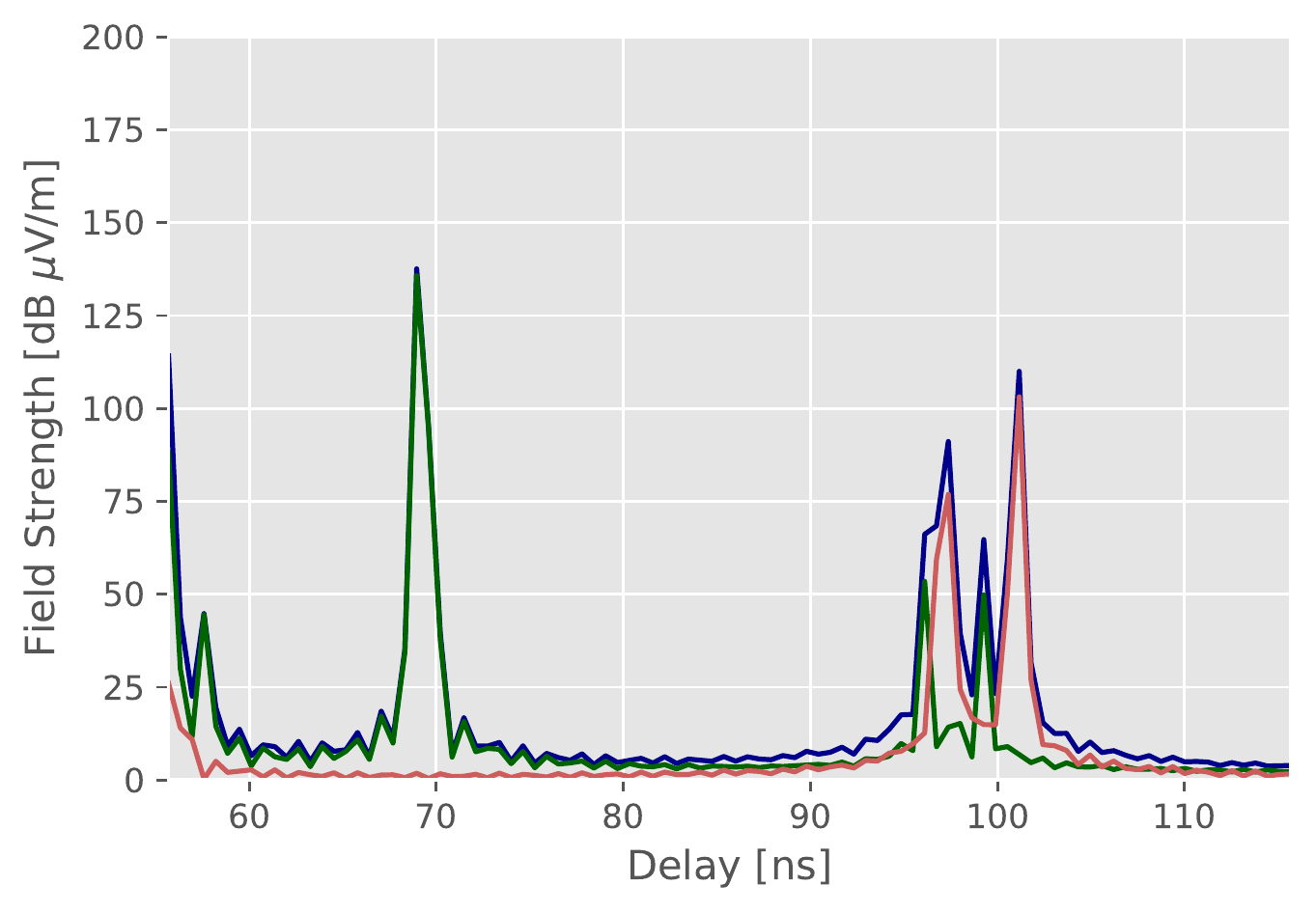}
    \label{fig:cir_substration}
}
\caption[]{Illustration of a generated CIR: \textbf{(a)} Propagation paths (pink) with reflections at the vehicle; \textbf{(b)} CIR generated by the radio propagation simulation with infinite bandwidth (red pulses) and sinc-interpolated bandlimited CIR (blue); \textbf{(c)} CIR subtraction (red) with vehicle (blue) and without vehicle (green).}
\label{fig:cir}
\end{figure*}

To ensure the correlation between the virtual simulation and actual environment, necessary outputs need to be unified. For multipath-assisted sensing, the CIR $h(t)$ represents the key input variable, as it houses all information of the multipath fading channel. Therefore, reflections at desired objects within the propagation environment are also included.

In the following, we will discuss the proposed approach given an in-house parking garage environment. Within this application, JC\&S systems can be applied to drastically increase the efficiency of limited parking capacities and the quality of user experience, by lowering parking space search time and costs. Specifically, benefits and multi-usage of integrated communication systems can provide cost-efficient and retrofittable technological solutions for digital parking management, including occupancy detection and digital billing.   

\subsection{Deterministic Radio Propagation Simulation}
\label{subsec:rps}
The propagation of radio signals is influenced by manifold propagation phenomena. This includes reflections, diffractions, or scattering caused by objects and obstacles within the environment. Especially reflections originating from objects cause measurable time delays and attenuation of the propagated signals. In order to emulate these signal paths, deterministic radio propagation simulation can calculate individual propagation paths with respect to the environment. This is achieved by approximating the emitted radio waves as rays \cite{Zhengqin_ray_tracing_2015}. In addition, various channel parameters and statistics between individual transceiver pairs, including path loss, received power, angular and delay spread, can be derived.

In general, the input of ray-tracing consists of the environmental model, which comprises a geometrical 3D model and associated material characteristics. In addition, a radio modal is required, which defines the physical layer properties of the desired radio technology, including center frequency, bandwidth, transmission power and additional parameters.

Given these inputs, ray-tracing deterministically calculates possible signal paths with respect to the spatial expansions, constellations and material specific parameters (e.g. relative permittivity and relative permeability) of objects. As a result, the length of each ray and therefore the delay at the receiver and the corresponding path loss is obtained.

\begin{figure}[ht]
    \centering
    \includegraphics[width=\linewidth]{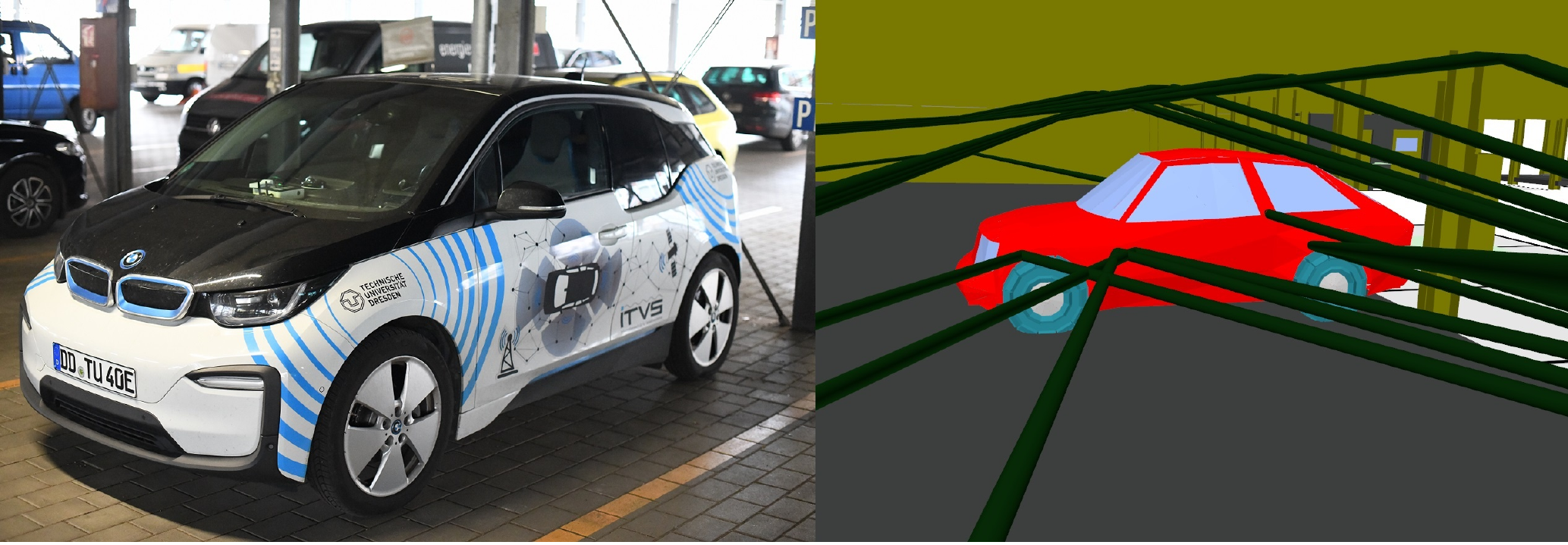}
    \caption{Exemplary illustration of a target vehicle (left) and a digital representative (right), including 3D propagation paths (dark green). Different material characteristics are shown in color.}
    \label{fig:3D_rays}
\end{figure}

For illustration purposes, a test vehicle and its virtual model within a parking garage environment is shown in Fig.~\ref{fig:3D_rays}. In addition, an exemplary ray-tracing between two transmitters, including reflections at the target vehicle, is shown. The scenario also corresponds to the ray-tracing results in Fig.~\ref{fig:cir}.

\subsection{CIR Reconstruction}
\label{subsec:reconstruction}
For multipath-assisted sensing, the CIR is a key input. In general, $h(t)$ consists of $k$ time-shifted impulses $\delta(t -\tau_k)$ with a specific amplitude $\alpha_k$ and time delay $\tau_k$:

\begin{equation}
\label{equ:CIR}
h(t) = \sum_{k=1}^{K} \alpha_k \delta (t-\tau_k).
\end{equation}

However, only a CIR given unlimited bandwidth is computed by the ray-tracing simulation. For a bandlimited channel, the pulse duration within the CIR corresponds to the available bandwidth and can be modulated by a normalized sinc function \cite{maymon_sinc_2011-1}. Thus, the Whittaker–Shannon interpolation formula or sinc interpolation allows the reconstruction of the bandlimited CIR $\overline{h}(t)$ from non-uniform samples $N$ following:  

\begin{equation}\label{equ:sinc}
   \overline{h}(t) = \sum_{n=0}^{N-1} \overline{h}(t_n) x(t-t_n) \quad \textrm{with }  x(t) = \sinc(\pi f_s t).
\end{equation}

For the reconstruction, the sum of normalized sinc functions for every sample in the CIR are used with the bandwidth of the bandlimited channel $f_s$. Fig.~\ref{fig:cir_bandlimited} shows an exemplary bandlimited CIR reconstruction (blue), given the ray-tracing results (red).

\begin{figure*}[t]
\centering
\subfigure[]{
    \includegraphics[trim=0 -60 0 0, clip, width=.41\textwidth]{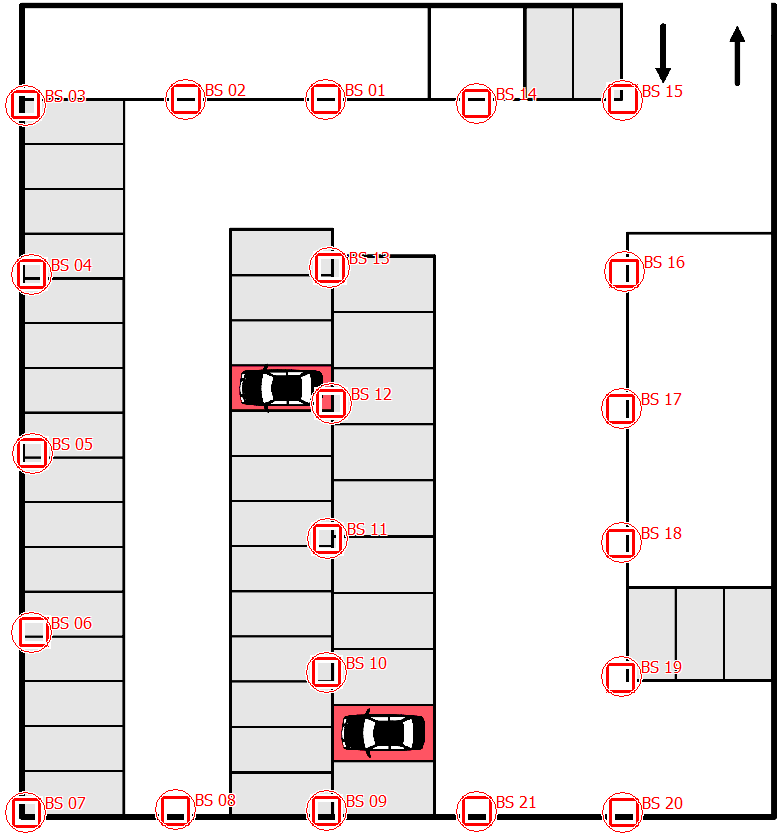}
    \label{fig:map_antenna}
}
\subfigure[]{
    \includegraphics[width=.55\textwidth]{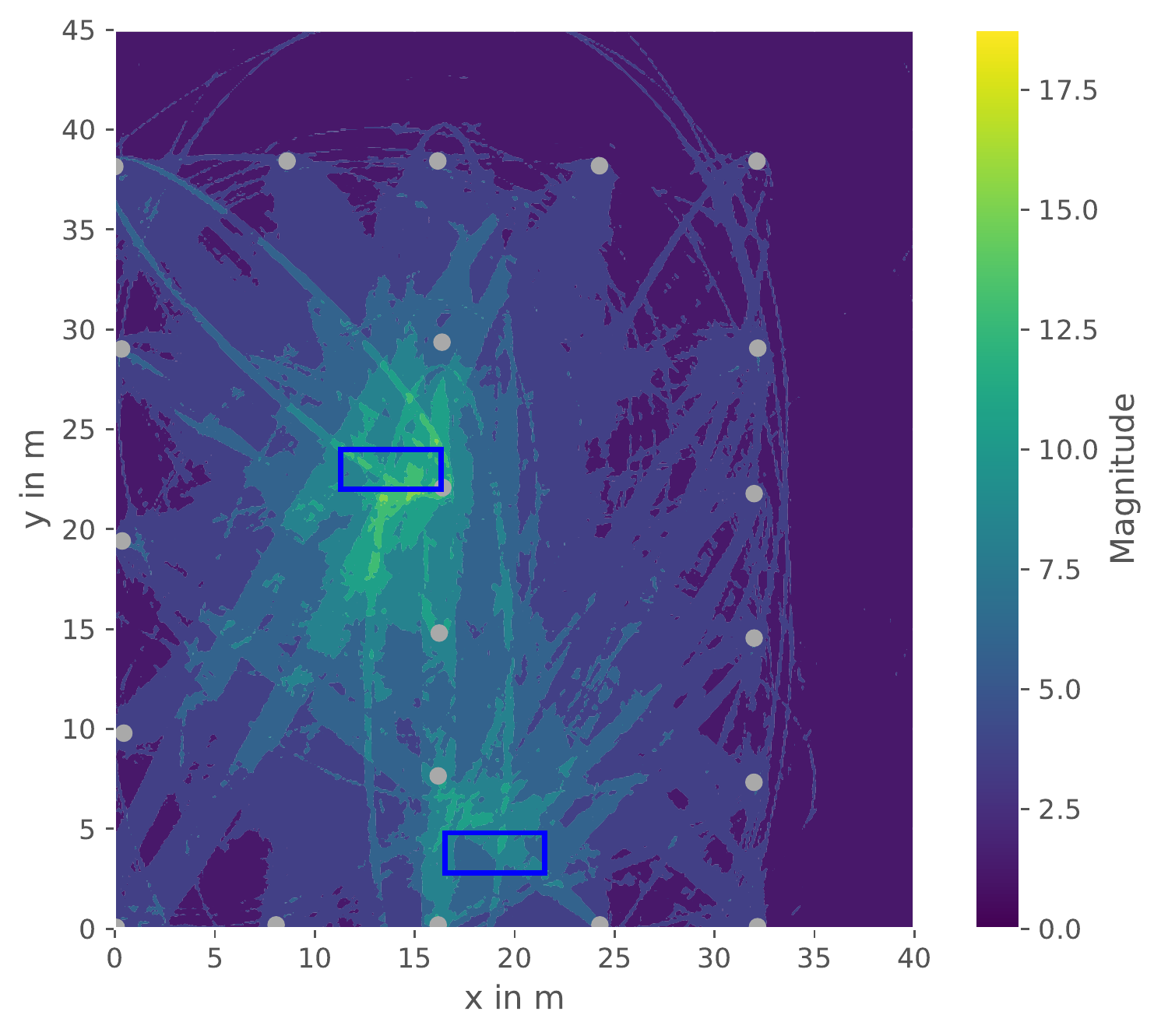}
    \label{fig:heatmap}
}
\caption[]{Network constellation and CIR mapping: \textbf{(a)} Base station placement in the parking garage and occupied parking lots; \textbf{(b)} Exemplary result of the heatmap sensing approach based on the subtracted CIR between all anchors (gray) with the reference position and expansions of the vehicles (blue box).}
\label{fig:map}
\end{figure*}

\subsection{Multipath-assisted Sensing}
\label{subsec:jcas_framework}
Complex real-world scenarios constitute spatially ambiguous reflection sources. Therefore, the distinction and mapping of reflecting objects is a challenging task for radio-based sensing systems. In \cite{ninnemann2022multipathassisted}, we proposed the application of a differential method, which compares the current observation with a previously surveyed reference CIR, representing a non-occupied scene. This background subtraction leads to a removal of the static environment and therefore only current reflection sources are represented in the resulting CIR. This procedure is illustrated in Fig.~\ref{fig:cir_substration}. The resulting CIR can then be used for mapping and positioning algorithms.

A general solution approach is the formulation of an elliptical model, as a transceiver pair essentially represents a bi-static constellation \cite{sensors-elliptical-model, sensors-elliptical-wall}. In order to reconstruct the environment, a CIR-based environmental mapping approach has been formulated in \cite{sensors_DFPL}. This method uses all interpolated and subtracted CIR values and estimates individual ellipses based on respective time delays, with the transceiver pair as focal points. In addition, the power values are also considered as they constitute to the magnitude of the time delays in the resulting map. This results in a weighted family of ellipses between transmitter and receiver, which is then interpolated and normalized across an equidistant grid, representing the state space. The sum of all normalized grids for every transmitter-receiver combination is a heatmap indicating the location and intensity of reflecting objects in the propagation environment. These calculation steps are then repeated for all available transceiver relations in order to create a representative map.

\section{Proof-of-Concept Evaluation}
\label{sec:Results}
In this section, the proposed method is empirically and qualitatively discussed, given the previously introduced in-house parking application. Here, the focus is on providing a proof-of-concept of the simulation setup. In the given scenario, 21 base stations (BS) are equally distributed throughout the parking area. In addition to sensing capabilities, this constellation can also be used for active localization functionalities \cite{Jung2020}. For illustration purposes, two vehicles (cf. Fig.~\ref{fig:3D_rays}), which are to be detected using radio sensing, are placed inside the area. The entire constellation is depicted in Fig.~\ref{fig:map_antenna}. The environment itself corresponds to the smart parking section in Fig.~\ref{fig:evaluation}.

As far as radio properties are concerned, we opted to emulate the physical layer characteristics of the 5G FR2 band \cite{etsi_ts_138_101_2}. The channels of the 5G FR2 feature a maximum bandwidth of \SI{400}{MHz}, which is not only beneficial for high data rate communication but also for higher range resolution in sensing applications. An overview of the parameters for the radio model of the propagation simulation is given in Tab.~\ref{tab:parameters}.

\begin{table}[h]
\centering
\caption{Radio Model - Simulation Parameter}
\label{tab:parameters}
    \begin{tabular}{@{}ll@{}}
        \toprule
        Parameter & Value \\ \midrule
        Center Frequency & \SI{26}{\giga \hertz} \\
        Bandwidth & \SI{400}{\mega \hertz} \\
        TX Power &  \SI{22}{\dBm}\\
        Antenna type & Omnidirectional antenna\\
        Number of BS & 21 \\
        Transceiver Height & \SI{1}{\meter} \\ \bottomrule
    \end{tabular}
\end{table}

Given the deterministic ray-tracing and bandlimited, reconstructed CIR, the ellipse-based environmental mapping surveillance method can be applied. A resulting heatmap is given Fig.~\ref{fig:heatmap}, where the vehicle positions are indicated by the blue boxes and the magnitude of the resulting heatmap corresponds to the reflection intensity of individual CIR samples. 

Even though, a parking lot selective sensing of the vehicles is not achieved, the resulting heatmap reveals higher intensities within the direct ambiance of the vehicles and therefore indicate reflection sources in those areas. In order to further improve the sensing resolution, a variety of approaches can be pursued. Due to the nature of the radio simulation, desired adaptations and modifications can easily be implemented and evaluated, inter alia including network constellations, target object amount and positions or radio properties, especially available bandwidth, or eventually sensing algorithms. 

\section{Conclusion \& Future work}
\label{sec:conclusion}
Radio sensing of future JC\&S systems will greatly enhance location-awareness and integrated surveillance capabilities of ITS. In this context, the paper included the discussion of potential multimodal ITS applications and their sensing accuracy demands. In order to provide a real-world proximate evaluation environment, especially for exploratory research and development, the usage of deterministic ray-tracing-based radio propagation is proposed. Based on pre-defined environments and testbeds, virtual models can be derived in order to represent the complexity of real-world scenarios.

The paper included a step-wise method for incorporating this scenario-based simulation into a possible radio sensing framework, aiming to provide real-world proximity, while also generating and using homogeneous parameters between possible measurements and the simulation platform. In order to do so, a toolchain aiming to integrate ray-tracing simulations for radio sensing applications, which use the derived CIR to generate an environmental heatmap, was presented and exemplary evaluated given an in-house parking scenario. This example demonstrated the capabilities of the presented simulation environment and provided a brief proof-of-concept of the presented sensing method. 


For future work, we will investigate the interrelationships between the real-world scenarios and digital representatives, in order to gain a deeper understanding of the requirements for radio sensing evaluation in complex environments and applications.

\bibliographystyle{IEEEtran}
\bibliography{lit.bib}

\begin{thebibliography}{10}
\providecommand{\url}[1]{#1}
\csname url@samestyle\endcsname
\providecommand{\newblock}{\relax}
\providecommand{\bibinfo}[2]{#2}
\providecommand{\BIBentrySTDinterwordspacing}{\spaceskip=0pt\relax}
\providecommand{\BIBentryALTinterwordstretchfactor}{4}
\providecommand{\BIBentryALTinterwordspacing}{\spaceskip=\fontdimen2\font plus
\BIBentryALTinterwordstretchfactor\fontdimen3\font minus
  \fontdimen4\font\relax}
\providecommand{\BIBforeignlanguage}[2]{{%
\expandafter\ifx\csname l@#1\endcsname\relax
\typeout{** WARNING: IEEEtran.bst: No hyphenation pattern has been}%
\typeout{** loaded for the language `#1'. Using the pattern for}%
\typeout{** the default language instead.}%
\else
\language=\csname l@#1\endcsname
\fi
#2}}
\providecommand{\BIBdecl}{\relax}
\BIBdecl

\bibitem{zhang_survey_data_driven_its_2011}
\BIBentryALTinterwordspacing
J.~Zhang, F.-Y. Wang, K.~Wang, W.-H. Lin, X.~Xu, and C.~Chen, ``Data-driven
  intelligent transportation systems: A survey,'' \emph{IEEE Transactions on
  Intelligent Transportation Systems}, vol.~12, no.~4, pp. 1624--1639, Dec.
  2011. [Online]. Available: \url{https://doi.org/10.1109/tits.2011.2158001}
\BIBentrySTDinterwordspacing

\bibitem{Papadimitratos_its_technologies_applications_outlook_2009}
\BIBentryALTinterwordspacing
P.~Papadimitratos, A.~L. Fortelle, K.~Evenssen, R.~Brignolo, and S.~Cosenza,
  ``Vehicular communication systems: Enabling technologies, applications, and
  future outlook on intelligent transportation,'' \emph{IEEE Communications
  Magazine}, vol.~47, no.~11, pp. 84--95, Nov. 2009. [Online]. Available:
  \url{https://doi.org/10.1109/mcom.2009.5307471}
\BIBentrySTDinterwordspacing

\bibitem{5G-ITS}
\BIBentryALTinterwordspacing
Z.~S. Bojkovic, D.~A. Milovanovic, and T.~P. Fowdur, Eds., \emph{5G Multimedia
  Communication}.\hskip 1em plus 0.5em minus 0.4em\relax {CRC} Press, Oct.
  2020. [Online]. Available: \url{https://doi.org/10.1201/9781003096450}
\BIBentrySTDinterwordspacing

\bibitem{bourdoux_6g_2020-3}
A.~Bourdoux, A.~N. Barreto, B.~van Liempd, C.~de~Lima, D.~Dardari, D.~Belot,
  E.-S. Lohan, G.~Seco-Granados, H.~Sarieddeen, H.~Wymeersch, J.~Suutala,
  J.~Saloranta, M.~Guillaud, M.~Isomursu, M.~Valkama, M.~R.~K. Aziz,
  R.~Berkvens, T.~Sanguanpuak, T.~Svensson, and Y.~Miao, ``6g white paper on
  localization and sensing,'' 2020.

\bibitem{ziegler_stratification_2019}
V.~Ziegler, T.~Wild, M.~Uusitalo, H.~Flinck, V.~R{\"a}is{\"a}nen, and
  K.~H{\"a}t{\"o}nen, ``Stratification of {{5G}} evolution and {{Beyond 5G}},''
  in \emph{2019 {{IEEE}} 2nd {{5G World Forum}} ({{5GWF}})}, Sep. 2019, pp.
  329--334.

\bibitem{aazhang_key_2019}
B.~Aazhang, P.~Ahokangas, H.~Alves, M.-S. Alouini, J.~Beek, H.~Benn, M.~Bennis,
  J.~Belfiore, E.~Strinati, F.~Chen, K.~Chang, F.~Clazzer, S.~Dizit, D.~Kwon,
  M.~Giordiani, W.~Haselmayr, J.~Haapola, E.~Hardouin, E.~Harjula, and P.~Zhu,
  \emph{Key Drivers and Research Challenges for {{6G}} Ubiquitous Wireless
  Intelligence (White Paper)}.\hskip 1em plus 0.5em minus 0.4em\relax 6G
  Flagship, Sep. 2019.

\bibitem{dang_what_2020-1}
S.~Dang, O.~Amin, B.~Shihada, and M.-S. Alouini, ``What should {{6G}} be?''
  \emph{Nature Electronics}, vol.~3, no.~1, pp. 20--29, Jan. 2020.

\bibitem{yang_6g_2019}
P.~Yang, Y.~Xiao, M.~Xiao, and S.~Li, ``{{6G Wireless Communications}}: Vision
  and {{Potential Techniques}},'' \emph{IEEE Network}, vol.~33, no.~4, pp.
  70--75, Jul. 2019.

\bibitem{zhang_6g_2019}
Z.~Zhang, Y.~Xiao, Z.~Ma, M.~Xiao, Z.~Ding, X.~Lei, G.~K. Karagiannidis, and
  P.~Fan, ``{{6G Wireless Networks}}: Vision, {{Requirements}},
  {{Architecture}}, and {{Key Technologies}},'' \emph{IEEE Vehicular Technology
  Magazine}, vol.~14, no.~3, pp. 28--41, Sep. 2019.

\bibitem{reid_localization_for_autonomous_vehicles_2019}
\BIBentryALTinterwordspacing
T.~G. Reid, S.~E. Houts, R.~Cammarata, G.~Mills, S.~Agarwal, A.~Vora, and
  G.~Pandey, ``Localization requirements for autonomous vehicles,'' \emph{SAE
  International Journal of Connected and Automated Vehicles}, vol.~2, no.~3,
  Sep. 2019. [Online]. Available: \url{https://doi.org/10.4271/12-02-03-0012}
\BIBentrySTDinterwordspacing

\bibitem{cao_joint_2020}
N.~Cao, Y.~Chen, X.~Gu, and W.~Feng, ``Joint {{Bi}}-{{Static Radar}} and
  {{Communications Designs}} for {{Intelligent Transportation}},'' \emph{IEEE
  Transactions on Vehicular Technology}, vol.~69, no.~11, pp. 13\,060--13\,071,
  Nov. 2020.

\bibitem{zhang_enabling_2021-1}
J.~A. Zhang, M.~L. Rahman, K.~Wu, X.~Huang, Y.~J. Guo, S.~Chen, and J.~Yuan,
  ``Enabling {{Joint Communication}} and {{Radar Sensing}} in {{Mobile
  Networks}} -{{A Survey}},'' \emph{IEEE Communications Surveys Tutorials}, pp.
  1--1, 2021.

\bibitem{ansari_DSRC_analysis_2017}
K.~Ansari, H.~S. Naghavi, Y.-C. Tian, and Y.~Feng, ``Requirements and
  complexity analysis of cross-layer design optimization for adaptive
  inter-vehicle dsrc,'' in \emph{International Conference on Mobile, Secure,
  and Programmable Networking}.\hskip 1em plus 0.5em minus 0.4em\relax
  Springer, 2017, pp. 122--137.

\bibitem{etsi_ts_122_261}
3GPP, ``Ts 122 261: 5g; service requirements for next generation new servies
  and markets,'' European Telecommunication Standards Institute (ETSI), Tech.
  Rep., 03 2019.

\bibitem{xinlie_urllc_use_cases_2019}
H.~Xinli and X.~Liang, ``Verticals urllc use cases and requirements,'' jul
  2019.

\bibitem{Cogalan_inflight_connectivity_2018}
\BIBentryALTinterwordspacing
T.~Cogalan, S.~Videv, and H.~Haas, ``Inflight connectivity: Deploying different
  communication networks inside an aircraft,'' in \emph{2018 IEEE 87th
  Vehicular Technology Conference (VTC Spring)}.\hskip 1em plus 0.5em minus
  0.4em\relax {IEEE}, Jun. 2018, pp. 1--6. [Online]. Available:
  \url{https://doi.org/10.1109/vtcspring.2018.8417707}
\BIBentrySTDinterwordspacing

\bibitem{Schultz_Cabin_OR_2019}
\BIBentryALTinterwordspacing
M.~Schultz and S.~Reitmann, ``Machine learning approach to predict aircraft
  boarding,'' \emph{Transportation Research Part C: Emerging Technologies},
  vol.~98, pp. 391--408, Jan. 2019. [Online]. Available:
  \url{https://doi.org/10.1016/j.trc.2018.09.007}
\BIBentrySTDinterwordspacing

\bibitem{Schwarzbach_covid_boarding_2020}
\BIBentryALTinterwordspacing
P.~Schwarzbach, J.~Engelbrecht, A.~Michler, M.~Schultz, and O.~Michler,
  ``Evaluation of technology-supported distance measuring to ensure safe
  aircraft boarding during {COVID}-19 pandemic,'' \emph{Sustainability},
  vol.~12, no.~20, p. 8724, Oct. 2020. [Online]. Available:
  \url{https://doi.org/10.3390/su12208724}
\BIBentrySTDinterwordspacing

\bibitem{liu_information-theoretic_2021}
\BIBentryALTinterwordspacing
Y.~Liu, M.~Li, A.~Liu, J.~Lu, and T.~X. Han, ``Information-{Theoretic} {Limits}
  of {Integrated} {Sensing} and {Communication} with {Correlated} {Sensing} and
  {Channel} {States},'' \emph{arXiv:2110.12683 [cs, math]}, Oct. 2021, arXiv:
  2110.12683. [Online]. Available: \url{http://arxiv.org/abs/2110.12683}
\BIBentrySTDinterwordspacing

\bibitem{etsi_tr_38_901}
3GPP, ``Tr 138 901: 5g; study on channel model for frequences from 0.5 to 100
  ghz,'' European Telecommunication Standards Institute (ETSI), Tech. Rep., 05
  2017.

\bibitem{sensors_DFPL}
J.~Ninnemann, P.~Schwarzbach, A.~Jung, and O.~Michler, ``Lab-based evaluation
  of device-free passive localization using multipath channel information,''
  \emph{Sensors}, vol.~21, no.~7, 2021.

\bibitem{mampi-uwbmultipath-assisted_2020-2}
M.~Cimdins, S.~O. Schmidt, and H.~Hellbr{\"u}ck,
  ``{{MAMPI}}-{{UWB}}\textemdash{{Multipath}}-{{Assisted Device}}-{{Free
  Localization}} with {{Magnitude}} and {{Phase Information}} with {{UWB
  Transceivers}},'' \emph{Sensors}, vol.~20, no.~24, p. 7090, Jan. 2020.

\bibitem{wild_joint_2021}
T.~Wild, V.~Braun, and H.~Viswanathan, ``Joint {{Design}} of {{Communication}}
  and {{Sensing}} for {{Beyond 5G}} and {{6G Systems}},'' \emph{IEEE Access},
  vol.~9, pp. 30\,845--30\,857, 2021.

\bibitem{Zhengqin_ray_tracing_2015}
\BIBentryALTinterwordspacing
Z.~Yun and M.~F. Iskander, ``Ray tracing for radio propagation modeling:
  Principles and applications,'' \emph{IEEE Access}, vol.~3, pp. 1089--1100,
  2015. [Online]. Available: \url{https://doi.org/10.1109/access.2015.2453991}
\BIBentrySTDinterwordspacing

\bibitem{ninnemann2022multipathassisted}
J.~Ninnemann, P.~Schwarzbach, and O.~Michler, ``Multipath-assisted radio
  sensing and occupancy detection for smart in-house parking in its,'' in
  \emph{International Conference on Indoor Positioning and Indoor Navigation},
  2021.

\bibitem{maymon_sinc_2011-1}
S.~Maymon and A.~V. Oppenheim, ``Sinc {{Interpolation}} of {{Nonuniform
  Samples}},'' \emph{IEEE Transactions on Signal Processing}, vol.~59, no.~10,
  pp. 4745--4758, Oct. 2011.

\bibitem{sensors-elliptical-model}
\BIBentryALTinterwordspacing
Q.~Lei, H.~Zhang, H.~Sun, and L.~Tang, ``A new elliptical model for device-free
  localization,'' \emph{Sensors}, vol.~16, no.~4, 2016. [Online]. Available:
  \url{https://www.mdpi.com/1424-8220/16/4/577}
\BIBentrySTDinterwordspacing

\bibitem{sensors-elliptical-wall}
\BIBentryALTinterwordspacing
D.~Kocur, M.~Švecová, and J.~Rovňáková, ``Through-the-wall localization of
  a moving target by two independent ultra wideband (uwb) radar systems,''
  \emph{Sensors}, vol.~13, no.~9, pp. 11\,969--11\,997, 2013. [Online].
  Available: \url{https://www.mdpi.com/1424-8220/13/9/11969}
\BIBentrySTDinterwordspacing

\bibitem{Jung2020}
\BIBentryALTinterwordspacing
A.~Jung, P.~Schwarzbach, and O.~Michler, ``Future parking applications:
  Wireless sensor network positioning for highly automated in-house parking,''
  in \emph{Proceedings of the 17th International Conference on Informatics in
  Control, Automation and Robotics}.\hskip 1em plus 0.5em minus 0.4em\relax
  {SCITEPRESS} - Science and Technology Publications, 2020. [Online].
  Available: \url{https://doi.org/10.5220/0009891107100717}
\BIBentrySTDinterwordspacing

\bibitem{etsi_ts_138_101_2}
3GPP, ``5g; nr; user equipment (ue) radio transmission and reception; part 2:
  Range 2 standalone (3gpp ts 38.101-2 version 15.2.0 release 15),'' European
  Telecommunication Standards Institute (ETSI), Tech. Rep. ETSI TS 138 101-2
  V15.2.0 (2018-07), 07 2018.

\end{thebibliography}

\end{document}